\documentclass[aps,preprint,nofootinbib,superscriptaddress,prc]{revtex4}
\usepackage{amssymb}
\usepackage{tipa}
\usepackage{epsfig}
\usepackage[bookmarksnumbered,bookmarksopen,colorlinks,citecolor=blue,linkcolor=blue]{hyperref}
\begin{document}

%\date{\today}

\title{Nucleus-nucleus potential from identical-particle interference}

\author{Ning Wang}
\email{wangning@gxnu.edu.cn}
\affiliation{ Department of Physics,
Guangxi Normal University, Guilin 541004, P. R. China}
\affiliation{Guangxi Key Laboratory of Nuclear Physics and Technology,
Guilin 541004, P. R. China}

\author{Yongxu Yang}
\affiliation{ Department of Physics,
Guangxi Normal University, Guilin 541004, P. R. China}
\affiliation{Guangxi Key Laboratory of Nuclear Physics and Technology,
Guilin 541004, P. R. China}

\author{Min Liu}
\affiliation{ Department of Physics,
Guangxi Normal University, Guilin 541004, P. R. China}
\affiliation{Guangxi Key Laboratory of Nuclear Physics and Technology,
Guilin 541004, P. R. China}

\author{Chengjian Lin}
\affiliation{ Department of Physics,
Guangxi Normal University, Guilin 541004, P. R. China}
\affiliation{China Institute of Atomic Energy, Beijing 102413, P. R. China}

\begin{abstract}
  Based on the quantum interference between two-identical-nucleus scattering at energies around the Coulomb barrier, the barrier positions for $^{58}$Ni+$^{58}$Ni and $^{16}$O+$^{16}$O are extracted from Mott oscillations in the angular distributions around 90$^{\circ}$ for the first time. The angle separation of pairs of Mott scattering valleys around 90$^{\circ}$ has a direct relationship with the closest distance between two nuclei in elastic scattering. Together with the barrier height from fusion excitation function, the extracted barrier position provides a sensitive probe to constrain the model predictions for the nucleus-nucleus potential barrier.

\end{abstract}

\maketitle

The study of nucleus-nucleus potential \cite{Bloc77,Myer00,Bass80,BW91,ETF4,Mis06,Lei95,New04,Zag02,Deni02,Gup05,Ada04} is one of the most important topics in nuclear physics. The basic features of the nucleus-nucleus potential are commonly described in terms of an interaction that is a function of the center-to-center distance between the projectile and target nuclei and consists of a repulsive Coulomb term and a short-ranged attractive nuclear component. The total potential possesses a maximum at a distance where the repulsive and attractive forces balance each other. This is referred to as the Coulomb barrier \cite{Das98,Bal98,Hag12,Back14}. For heavy-ion fusion and scattering reactions, the Coulomb barrier directly influences the behavior of fusion and scattering cross sections, and an accurate extraction of the barrier not only the height but also the position and curvature from fusion cross sections and angular distributions of elastic/inelastic scattering attracted therefore a lot of attentions in many decades. Based on some empirical or realistic nuclear interactions \cite{Bloc77,Myer00,Bass80,BW91,ETF4,Mis06} together with barrier penetration concept \cite{Wong73,Qin12}, coupled-channel methods \cite{Hag99,Beck07}, optical models \cite{Daeh80,Yang04,Li07} or microscopic dynamics equations \cite{Wash08,Guo07,Ober14,ImQMD2014,SmGd}, one attends to obtain the information of the nucleus-nucleus potential barrier through reproducing the measured cross sections. Unfortunately, it is found that the data can be reproduced reasonably well with different potentials combining different theoretical models \cite{Daeh80,Zhou17}. Because of the ambiguity of optical model potential (commonly known as the Igo ambiguity \cite{Igo59}) due to the complicated parameter space and internal structure of the reaction partners, it is therefore of great significance to directly extract the barrier from measured data or to reduce the model dependence as much as possible. The height of the Coulomb barrier (or the distribution of the barrier height) could be extracted from the precisely measured fusion cross sections \cite{liumin06,Wang09,Gup92,Wangbin} or the back-angle quasi-elastic scattering cross sections \cite{Timm97,Zhang08} through a simple and elegant mathematical transformation. Comparing with the barrier height, the extraction of the positions of the Coulomb barrier in heavy-ion reactions at energies around the Coulomb barrier has not yet been explored extensively. Conventionally, the barrier position $R_B$ may be extracted from the classical formula $\sigma_{\rm fus}=\pi R_B^2(1-V_B/E_{c.m.})$ for fusion reactions at energies above the barrier height $V_B$ \cite{Birk78}. The extracted result with this conventional method which is model dependent with assumption that all $l$ waves contributing to the fusion cross sections have the same barrier position $R_B$ \cite{Birk78}, is sensitive to the selected fusion cross sections in the analysis and the uncertainty is relatively large. For example, with a linear fit of the measured fusion cross sections at above barrier energies for $^{16}$O+$^{16}$O \cite{Thom86}, one obtains a value of $R_B=8.7\pm 0.5$ fm, whereas a value of $R_B=9.8 \pm 0.6$ fm is obtained with the data from Ref. \cite{Wu84} for the same reaction. It is therefore required to further investigate the barrier position from heavy-ion fusion/scattering cross sections.

\begin{figure}
\includegraphics[angle=-0,width=0.6\textwidth]{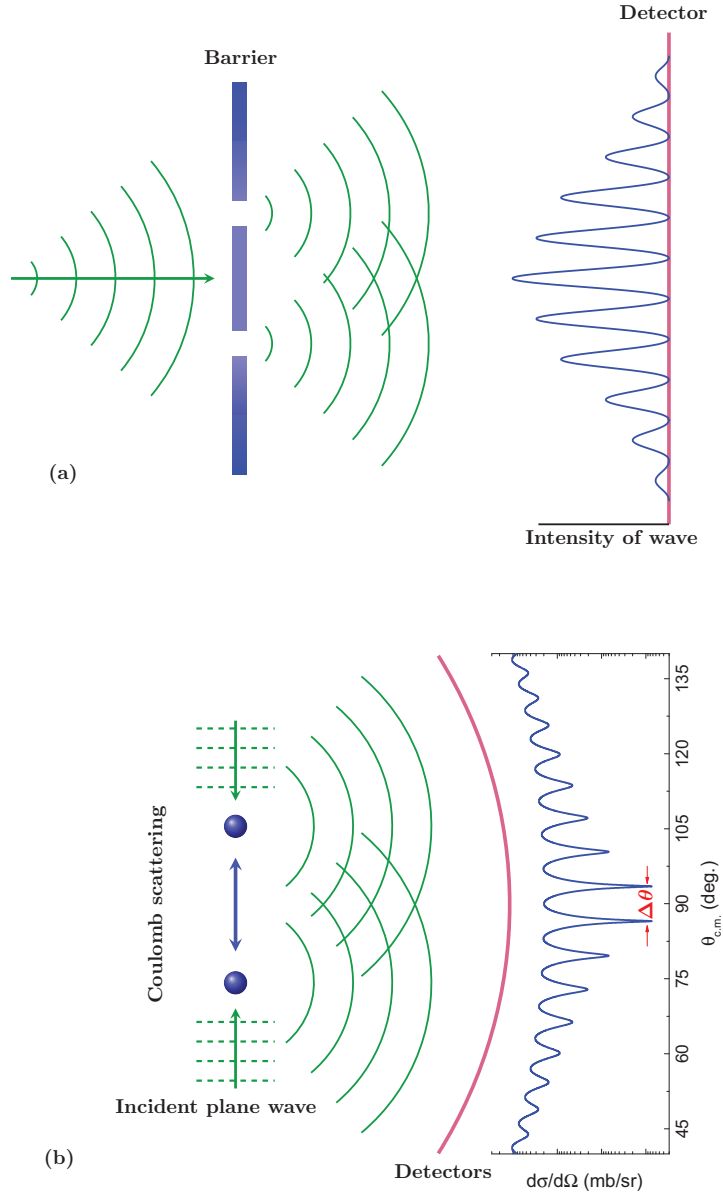}
 \caption{ (Color online) (a) Schematic view of the double slit interference of light. (b) Schematic view of a Mott scattering between two identical nuclei.}
\end{figure}

For microscopic particles, such as photons, nucleons, heavy nuclei and so on, the quantum effect especially quantum interference between two-identical-particle have been evidently observed, which verifies the wave properties of microscopic particles, such as the positions of the sources of waves and de Broglie wavelength. In the double slit interference experiment of light as shown in Fig.1(a), the slit separation $d$ has a relationship with the linear separation $\Delta y$ between fringes on screen (detector):
\begin{equation}
d  \simeq \lambda L/\Delta y
\end{equation}
where $\lambda$ is the wavelength of light, $L$ is the distance between the slit and the screen. It is known that the slit separation should be comparable with the wavelength of light in order to form the evident fringes on the screen. Although the direct double slit interference experiment for heavy nuclei is difficult since the slit separation should be at the femtometer scale, the quantum interference was clearly observed from the angular distribution of elastic scattering between identical projectile and target nuclei as shown in Fig.1(b), which is known as the Mott scattering \cite{Mott,Brom61,Hind07}.
Mott proposed an analytical formula for describing the differential cross section in the center-of-mass system for pure Coulomb scattering of identical particles \cite{Mott}:
\begin{eqnarray}
d \sigma/d\Omega &=& \frac{Z^4e^4}{16E_{c.m.}^2}   \{ \csc^4(\frac{\theta_{c.m.}}{2})+\sec^4(\frac{\theta_{c.m.}}{2})
\nonumber \\
&+& 2\frac{(-1)^{2I}}{2I+1}\csc^2(\frac{\theta_{c.m.}}{2})\sec^2(\frac{\theta_{c.m.}}{2})\cos[ \eta \ln (\tan^2(\frac{\theta_{c.m.}}{2}))]   \},
\end{eqnarray}
where $Z$ is the charge number of nuclei, $\theta_{c.m.}$ is the center-of-mass scattering angle, $E_{c.m.}$ is the center-of-mass energy and $I$ is the intrinsic spin of particles.
$\eta=\frac{Z^2e^2}{\hbar v}$ is the Sommerfeld number which is  $\frac{1}{2}$ the ratio of
the characteristic distance of closest approach given by $Z^2e^2/E_{c.m.}$ and the reduced wavelength $\lambdabar$ \cite{Brom61}.
According to Eq.(2), the Mott oscillations can be seen most clearly around $90^\circ$. The separation $\Delta \theta$ around $90^\circ$ could have a direct relationship with the closest distance between the two nuclei in scattering.

We first systematically investigate the Mott oscillations for a series of elastic scattering reactions at energies below the Coulomb barrier. From light to heavy systems, the separation $\Delta \theta$ of Mott oscillations around $90^\circ$ [see Fig.1(b)] decreases with the masses of nuclei due to the decrease of the corresponding de Broglie wavelength $\lambda=h/\sqrt{2\mu E_{c.m.}}$, where $\mu$ is the reduced mass of the system. Similar to Eq.(1), we note that the closest distance $d$ between two heavy nuclei in pure Coulomb scattering approximately satisfies $d \approx \lambda/ \sin(\Delta \theta)$. In Fig. 2, we show the discrepancies betweem the calculated closest distance with the separation $\Delta \theta$  and the expected value $d=Z^2 e^2/E_{c.m.}$. One can see that the closest distances are described very well with $d \approx \lambda/ \sin(\Delta \theta)$ for heavy nuclei. However, this approximation is not good enough for light systems, and the discrepancies increase rapidly with the value of $\Delta \theta$ and even larger than 0.35 fm for $^{16}$O+$^{16}$O. To improve the accuracy, we propose a modified expression:
\begin{equation}
d  \simeq   \frac{2 \lambda}{\sin(\Delta \theta)+\tan(\Delta \theta)}.
\end{equation}
From Fig. 2, one can see that for all investigated systems the accuracy of the calculated closest distances with Eq.(3) are significantly improved. Experimentally, the Mott oscillations of $^{58}$Ni+$^{58}$Ni at an incident energy of $E_{c.m.}=80$ MeV which is lower than the Coulomb barrier by about 20 MeV, were precisely measured in Ref. \cite{Hind07} and the separation of Mott oscillations around $90^\circ$  is $\Delta \theta=2.42 ^\circ \pm 0.03^\circ$. The obtained closest distance according to Eq.(3) is $14.12  \pm  0.17$ fm which is consistent with the expected value of 14.11 fm from $d=Z^2 e^2/E_{c.m.}$.

 \begin{figure}
\includegraphics[angle=-0,width=0.8\textwidth]{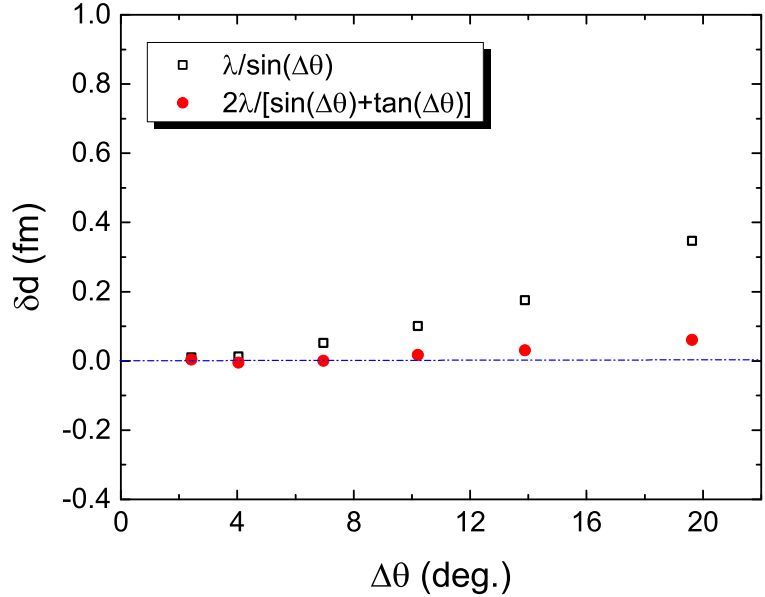}
 \caption{(Color online)  Discrepancies between the calculated closest distance based on the separation of Mott oscillations around $90^\circ$ and the expected value $d=Z^2 e^2/E_{c.m.}$ for a series of reactions at sub-barrier Coulomb scattering. }
\end{figure}

 \begin{figure}
\includegraphics[angle=-0,width=0.75\textwidth]{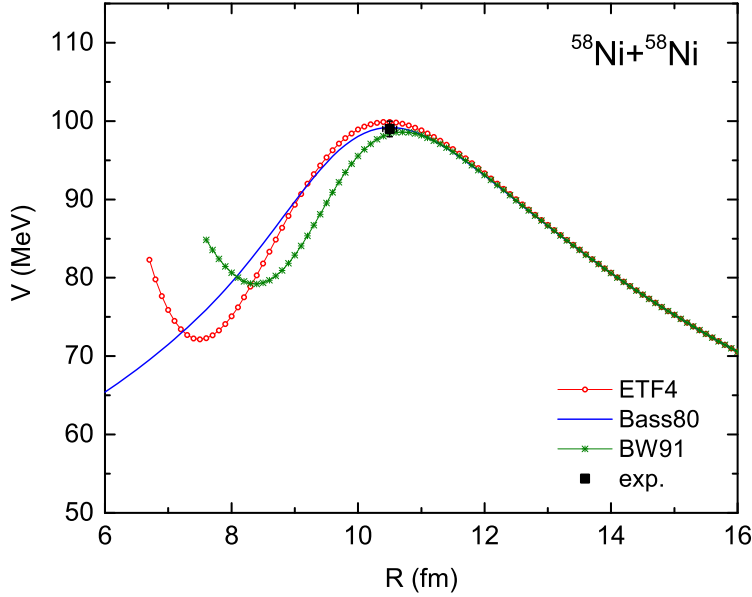}
 \caption{(Color online)  Nucleus-nucleus potential for $^{58}$Ni+$^{58}$Ni. The curves denote three empirical potentials and the square denotes the extracted barrier by using Eq.(3) together with the barrier height from the fusion excitation function \cite{Beck81}.}
\end{figure}

With Eq.(3), we further investigate the Coulomb barrier of $^{58}$Ni+$^{58}$Ni. The Mott oscillations around $90^\circ$ in the quasi-elastic scattering were also measured at an incident energy of $E_{c.m.}=100$ MeV \cite{Hind07} which is generally at the Coulomb barrier. Considering that the elastic cross sections at energies around the Coulomb barrier are dominant in the quasi-elastic scattering and the adding of the inelastic cross sections does not change significantly the angle separation around $90^\circ$ according to the coupled-channel calculations \cite{Hag99}, the angle separation in elastic scattering of $^{58}$Ni+$^{58}$Ni is directly analyzed based on the measured oscillations for quasi-elastic scattering. The obtained separation $\Delta \theta =2.91 ^\circ \pm 0.03^\circ$ which is slightly larger than the prediction of Mott's formula due to the influence of nuclear force. The corresponding closest distance according to Eq.(3) is $d=10.49\pm 0.11$ fm. In Fig. 3, we show the nucleus-nucleus potential for $^{58}$Ni+$^{58}$Ni. The solid curves, the circles and  the crosses denote the empirical Bass potential \cite{Bass80}, Broglia-Winther (BW) potential \cite{BW91} and the potential based on extended Thomas-Fermi (ETF) approximation \cite{ETF4}, respectively. The square denotes the extracted Coulomb barrier. One sees that all of the three potentials are in good agreement with the extracted result at the top of the potential.

 \begin{figure}
\includegraphics[angle=-0,width=1.0\textwidth]{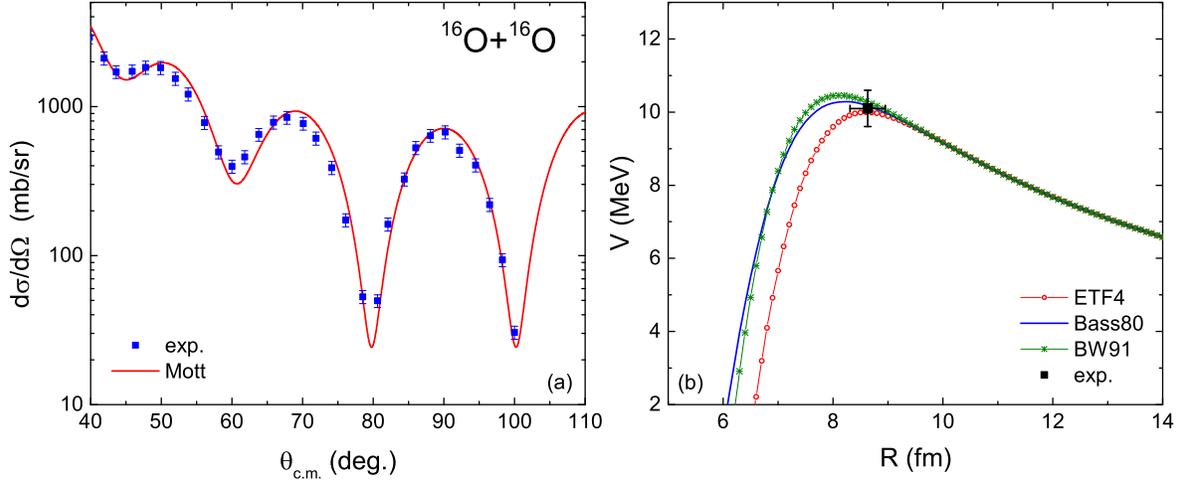}
 \caption{(Color online) (a) Elastic scattering angular distribution for $^{16}$O+$^{16}$O at $E_{c.m.}=10.91$ MeV. The curve denotes the results of Mott's formula. (b) The same as Fig. 3, but for $^{16}$O+$^{16}$O.   }
\end{figure}

In addition, the Coulomb barrier of $^{16}$O+$^{16}$O is also investigated. The fusion excitation functions of $^{16}$O+$^{16}$O have been measured by several groups \cite{Kola79,Wu84,Thom86,Kuro87} and the extracted mean barrier height is $10.1\pm 0.5$ MeV. Simultaneously, the elastic scattering angular distributions around $90^\circ$ were also measured at an energy of 10.91 MeV in Ref. \cite{Wu84} and the data are shown in Fig. 4(a). The Mott oscillations can be clearly observed around $90^\circ$. The data are reproduced roughly well with the Mott's formula which is based on pure Coulomb scattering. Through Gaussian fits to the data we obtain the separation of the Mott oscillations $\Delta \theta =21.04 ^\circ \pm 0.80^\circ$. The corresponding closest distance is $d=8.63\pm0.32$ fm. We note that the uncertainty of the extracted barrier position with the proposed method is significantly reduced comparing with that from the conventional method mentioned previously. In Fig. 4(b), we compare the extracted Coulomb barrier for $^{16}$O+$^{16}$O and the predictions from the three empirical potentials. It looks that the ETF4 potential reproduces the extracted barrier better. We also note that the predicted height and position of the Coulomb barrier for $^{16}$O+$^{16}$O are $V_B=10.12$ MeV and $R_B=8.52$ fm from the time-dependent Hartree-Fock (TDHF) theory \cite{Wash08}, which are in good agreement with the extracted values.

In the proposed method, one needs to know the barrier height beforehand and then to measure the angle separation in Mott oscillations around $90^\circ$ at energies around the barrier. It is therefore necessary to check the sensitivity of the extracted values of barrier position with the change of barrier height. The uncertainty of the extracted barrier height based on a precisely measured fusion excitation function is generally smaller than one MeV. Through adopting different optical potentials, we vary the barrier height within one MeV (but remain the position of potential barrier fixed) and find that the calculated change of the separation $\Delta \theta$ is smaller than $1\%$ for both $^{58}$Ni+$^{58}$Ni and $^{16}$O+$^{16}$O at energies around the barrier. We also note that the measured $\Delta \theta$ for $^{16}$O+$^{16}$O at $E_{c.m.}=11.92$ MeV is almost the same as that at 10.91 MeV \cite{Wu84}. In addition, it is known that at energies around the Coulomb barrier, the channel coupling effects lead to a distribution of fusion barrier, with which one obtains a mean value of the barrier height. Based on the elastic scattering between two-identical nuclei, the obtained barrier position could be a mean value of the positions of potential barriers considering the channel coupling effects.

To summarized, the quantum interference between two-identical-particle can be clearly observed not only for photons but also for heavy nuclei. Inspired by the double slit interference experiment of light, the separation of the Mott oscillations around 90$^{\circ}$ in the elastic scattering angular distributions provides a sensitive model-independent probe to investigate the position of the Coulomb barrier. By using an analytical formula with high accuracy, the Coulomb barriers for $^{58}$Ni+$^{58}$Ni and $^{16}$O+$^{16}$O are extracted and compared simultaneously with some model predictions.

\begin{center}
\textbf{ACKNOWLEDGEMENTS}
\end{center}

This work was supported by National Natural Science Foundation of
China (Nos. 11422548, 11635015, 11365005, 11365004 and 11747307), Guangxi Natural Science Foundation (No. 2015GXNSFDA139004), and the Foundation of Guangxi innovative team and distinguished scholar in institutions of higher education.   We thank Xiaohua Li for helpful communications and Li Ou for valuable discussions.

\end{document}